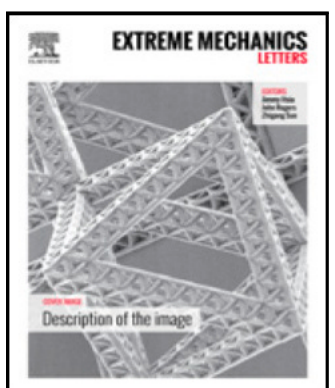

# Tensile wrinkling and creasing of an elastic half-space under a suction load

Roberta Springhetti, Davide Bigoni *

*Instabilities Lab, University of Trento, via Mesiano 77, Trento, Italy*

ABSTRACT

A famous and thoroughly investigated instability set-up, susceptible to wrinkling and creasing, consists of an elastic half-space being prestressed under a *dead load*, applied at infinity and, possibly, on its surface. We consider the case of a *pressure* or a *suction* applied on the surface and show that wrinkling and creasing occur in a surprising way, completely different from dead load. With reference to incompressible neo-Hookean elasticity and assuming the prevalence of a uniaxial prestress state induced by the application of a uniform pressure, no wrinkling or creasing is foreseen, whereas – unexpectedly – they do when reversing the sign of pressure, which is then a suction, thus leading to *tensile creasing* and *tensile wrinkling*. When a biaxial prestress state is considered, it is shown that some loading paths can be envisaged, able to grow to infinity without causing any instability. Consequently, we suggest that, according to its sign, pressure can promote or delay surface instabilities, a finding which may have implications for mechanobiology or microfluidic fluid–structure interaction.

## 1. Introduction

More than sixty years ago, Biot [1,2] discovered that an elastic half-space subject to a sufficiently intense compression parallel to its free surface may bifurcate into a mode exhibiting a diffuse undulation exponentially decaying with depth, a phenomenon often referred to as 'surface bifurcation' or 'wrinkling'. Since then, this problem has been the subject of extensive theoretical exploration, see among many [3–7]. Experimental investigations [8,9] reveal that the Biot's surface bifurcation is generally preceded by the so-called 'creasing', an instability characterized by the appearance of a surface folding, which evolves into a self-contacting sulcus as the load increases. On the basis of a numerical investigation, Hohlfeld and Mahadevan [10] recognized creasing as a scale-free subcritical nonlinear instability mode, distinct from the surface instability investigated by Biot. This is consistent with the interpretation suggested by Cao and Hutchinson [11] that wrinkling represents a bifurcation extremely sensitive to imperfections. According to this approach, the free surface becomes metastable for a prestress lying between the thresholds for creasing and wrinkling, so that the flat configuration turns out to be linearly stable, but nonlinearly unstable, see also [12,13]. This view is the subject of the discussion in [14]. However, our aim is not to contribute to this debate, but rather to show that wrinkling and creasing can occur under conditions that have so far gone unnoticed. In particular, in all the above quoted works, the half-space is subject to a remotely applied *dead loading*, leaving its surface free. Conversely, when a dead load is added onto the surface, a biaxial prestress develops with components parallel and normal to it (the out-of-plane component is not considered). Wrinkling is well-known under these conditions [4], while creasing has not been investigated so far. Although the analysis in the presence of a dead load on the surface is not our objective, it is included here as a reference for comparison. Rather, the present article focuses on the effects of a uniform pressure or suction acting on the surface of the half-space.

Within the standard framework of incompressible neo-Hookean elasticity in the context of two-dimensional deformation, the analyses of wrinkling (based on a linearized bifurcation) and creasing (carried out via finite elements according to the procedure suggested by Suo and co-workers [15]) lead to the results summarized in Table 1 for a half-space under a prestress normal to its surface in the elliptic regime.

In particular, when the half-space is subjected to surface *pressure* or *suction*, the following features emerge.

(*i*) The mechanical conditions required for the formation of creases and wrinkles are completely different from those involved in the well-known case of dead load. In particular, the region where wrinkling is excluded turns out to be unbounded, so that there exist loading paths along which the prestress may increase without bound, while preserving stability.

(*ii*) Surprisingly, when the load is a *pressure* inducing a uniaxial compressive state, the half-space does not become unstable to wrinkling or creasing. However, both instabilities become possible when the sign of pressure is reversed. In other words, an elastic half-space exhibits creasing, followed by wrinkling at a higher load intensity, when a uniform suction is applied along its surface, producing a uniaxial tensile stress perpendicular to it.

* Corresponding author.
*E-mail address:* bigoni@ing.unitn.it (D. Bigoni).

The above results are complemented by an analysis of wrinkling for anisotropic elasticity, showing that the main features found for pressure and suction persist in a context more general than neo-Hookean elasticity. This means that tensile bifurcation is not merely an ‘accident’ arising from a particular choice of constitutive parameters. Moreover, our dead load analyses, carried out for comparison, revealed a complementary, yet unexpected result: although wrinkling develops as effect of a compressive load applied on the surface, *creasing is absent*. Therefore, the symmetry of the instability with respect to the two principal stresses generated via dead load is lost.

Pressure and suction loads are becoming increasingly relevant in light of their wide range of applications in mechanobiology (where cilia, flagella, or lower level organisms interact with fluids for motility or feeding purposes [16]), or in micro–fluidics (for soft actuation [17,18]), or in the mechanics of growth [19]. Therefore, the results reported in the present article suggest possible applications to fluid–structure interaction involving motility, or soft actuation, or morphogenesis.

## 2. Bifurcation of an incompressible elastic half-space under pressure/suction and dead load

Referring to the kinematics of a solid, the deformation gradient $\mathbf{F}$ and its determinant $J$ are introduced, so that the Nanson’s rule for the transformation of an oriented area element $\mathbf{n}_0 \mathrm{d}a_0$, as defined in the reference configuration, into the corresponding element $\mathbf{n}\mathrm{d}a$, mapped in the current configuration, is

$$\mathbf{n}\,\mathrm{d}a = J\mathbf{F}^{-T}\mathbf{n}_0\,\mathrm{d}a_0; \tag{1}$$

this relation leads to the definition of the nominal stress $\mathbf{t}$ (the transpose of the Piola stress) as related to the Cauchy stress $\mathbf{T}$ through

$$\mathbf{T}\mathbf{n}\,\mathrm{d}a = \mathbf{t}^T\mathbf{n}_0\,\mathrm{d}a_0. \tag{2}$$

### 2.1. Loads on the surface of a half-space

Two types of loading are considered to be applied along the surface of an elastic half-space. In addition to the pressure/suction load, which is the focus of the present study, dead loads, uniform and periodically distributed, are also considered, the latter acting as perturbing agent in the sense proposed by Biot [1].

- **Pressure/suction load** A *pressure load* $p$ (assumed positive when compressive) applied to a surface of unit normal $\mathbf{n}$ in the current configuration prescribes that the Cauchy traction $\mathbf{T}\mathbf{n}$, but also the nominal (or Piola) traction $\mathbf{t}^T\mathbf{n}_0$ remain aligned with $\mathbf{n}$,

$$\mathbf{T}\mathbf{n}\,\mathrm{d}a = -p\mathbf{n}\,\mathrm{d}a = \mathbf{t}^T\mathbf{n}_0\,\mathrm{d}a_0 = -pJ\mathbf{F}^{-T}\mathbf{n}_0\,\mathrm{d}a_0, \tag{3}$$

  where $\mathrm{d}a$ and $\mathrm{d}a_0$ are the area elements in the current and reference configurations, respectively. *Suction* is defined here as a negative, namely *tensile* pressure, and thus the orientation of the load is determined by the sign of $p$, which is negative for suction and positive for pressure. A material differentiation with respect to a time-like parameter governing the deformation of Eq. (3) yields the incremental condition

$$\dot{\mathbf{t}}^T\mathbf{n}_0 = -J\left[\dot{p}\mathbf{I} + p(\mathrm{tr}\mathbf{L})\mathbf{I} - p\mathbf{L}^T\right]\mathbf{F}^{-T}\mathbf{n}_0, \tag{4}$$

  where $\mathbf{L} = \dot{\mathbf{F}}\mathbf{F}^{-1}$ is the gradient of incremental displacement. For the elastic half-space $x_1 \geq 0$, having boundary outward unit normal $\mathbf{n}_0 = -\mathbf{e}_1$, the incremental boundary conditions (4) in component form become in an updated Lagrangean deformation where the current configuration is assumed as reference

$$\dot{t}_{11} = -\dot{p} + p\,v_{1,1}, \qquad \dot{t}_{12} = p\,v_{1,2}, \tag{5}$$

  where $v_i$ $(i = 1,2)$ are the incremental displacements, so that $L_{ij} = v_{i,j}$, and the incompressibility constraint, $J = 1$, becomes $\mathrm{tr}\mathbf{L} = v_{i,i} = 0$.
- **Dead load** A *dead load* $\mathbf{d}$, defined in the reference configuration, and whose direction remains parallel to the reference unit normal $\mathbf{n}_0$,

$$\mathbf{t}^T\mathbf{n}_0 = d\mathbf{n}_0, \tag{6}$$

  is considered to prescribe the assumed initial stress state in the half-space. As $\mathbf{n}_0 = -\mathbf{e}_1$, the boundary conditions for the incremental problem become

$$\dot{t}_{11} = \dot{d}, \qquad \dot{t}_{12} = 0. \tag{7}$$

  At this point, it is instrumental to introduce a variable dead load increment to be used in the subsequent analysis of wrinkling following Biot [1]. The sinusoidal load, having amplitude $\sigma$ and wavelength $\ell$

$$\dot{q}(x_2) = \sigma\cos(\zeta x_2), \qquad \zeta = m\pi/\ell, \qquad m = 1,2,3,\ldots \tag{8}$$

  is added on the half-space surface $x_1 = 0$ to the load (dead, pressure or suction) generating the prestress.

Once the prescribed state of prestress is enforced, the periodic incremental dead load (8) is superimposed to perturb the underlying homogeneous solution. As $\dot{p} = \dot{d} = 0$, all loads can be described as

$$\dot{t}_{11} = \dot{q}(x_2) + \mu\alpha\,v_{1,1}, \qquad \dot{t}_{12} = \mu\alpha\,v_{1,2}, \tag{9}$$

where $\mu$ denotes the incremental shear modulus, as defined in the following Section, and

$$\alpha = p/\mu, \quad \text{for pressure/suction loading,}$$
$$\alpha = 0, \quad \text{for dead loading.}$$

### 2.2. Incremental constitutive equations

An incompressible and initially orthotropic material, experiencing deformation under a predominant plane strain assumption, responds to an in-plane incremental deformation as [4]

$$\begin{aligned} \dot{t}_{11}/\mu &= (2\xi - k - \eta)\,v_{1,1} + \dot{\Pi}/\mu, & \dot{t}_{12}/\mu &= (1+k)v_{2,1} + (1-\eta)v_{1,2}, \\ \dot{t}_{21}/\mu &= (1-\eta)v_{2,1} + (1-k)v_{1,2}, & \dot{t}_{22}/\mu &= (2\xi + k - \eta)\,v_{2,2} + \dot{\Pi}/\mu, \end{aligned} \tag{10}$$

where $\dot{\Pi}$ represents the increment of the mean in-plane stress $\Pi$ – the Lagrangian multiplier enforcing the incompressibility constraint – whereas parameters

$$\xi = \frac{\mu_*}{\mu}, \qquad k = \frac{T_1 - T_2}{2\mu}, \qquad \eta = \frac{\Pi}{\mu} = \frac{T_1 + T_2}{2\mu} \tag{11}$$

account for the effects of prestress through the in–plane principal Cauchy stresses $T_1$ and $T_2$, and the incremental shear moduli $\mu_*$ and $\mu$.

**Table 1**
Critical prestress $T_1$ for creasing and wrinkling, normalized by the incremental shear modulus $\mu$ for a neo-Hookean half-space under a uniform surface load inducing $T_1$ with $T_2 = 0$.

| Load | Compression | | Tension | |
|---|---|---|---|---|
| | Wrinkling | Creasing | Creasing | Wrinkling |
| Pressure | – | – | $T_1/\mu = +1.41$ | $T_1/\mu = +1.68$ |
| Dead | $T_1/\mu = -1.68$ | – | – | – |

The elliptic (complex and imaginary) regime is defined by the following constitutive restrictions:

$$\mu > 0, \qquad k^2 < 1, \qquad 2\xi > 1 - \sqrt{1-k^2}. \tag{12}$$

Under a predominant in-plane uniaxial prestress $T_1$ ($T_2 = 0$) normal to the surface of the half-space, $\eta = k = T_1/(2\mu)$, so that within the elliptic regime, a uniaxial prestress is restricted to $-2\mu < T_1 < 2\mu$.

### 2.3. The incremental deformation of an elastic incompressible solid

In the absence of body forces, the incremental nominal stress components $\dot{t}_{ij}$ obey the equilibrium equations, $\dot{t}_{ij,i} = 0$, so that, upon substitution of the nominal stress components, Eq. (10), and use of the incompressibility constraint, the gradient of the in-plane mean stress turns out to be

$$\begin{aligned} \dot{\Pi}_{,1} &= \mu\left[(1+k-2\xi)\,v_{1,11} - (1-k)v_{1,22}\right], \\ \dot{\Pi}_{,2} &= \mu\left[-(1+k)v_{2,11} + (1-k-2\xi)\,v_{2,22}\right]. \end{aligned} \tag{13}$$

For an incompressible solid under incremental plane strain, a stream function $\psi(x_1, x_2)$ is introduced to define the incremental displacements as

$$v_1 = \psi_{,2}, \quad v_2 = -\psi_{,1}, \tag{14}$$

so that incompressibility is automatically satisfied. The elimination of $\dot{\Pi}$ from Eqs. (13) allows to obtain the following condition for the stream function

$$(1+k)\,\psi_{,1111} - 2\,(1-2\xi)\,\psi_{,1122} + (1-k)\,\psi_{,2222} = 0. \tag{15}$$

A separate-variables solution for Eq. (15) can be sought in the form [4]

$$\psi(x_1,x_2) = -\frac{1}{i\kappa}\left[b_1 e^{i\kappa\Omega_1 x_1} + b_2 e^{i\kappa\Omega_2 x_1} + b_3 e^{i\kappa\Omega_3 x_1} + b_4 e^{i\kappa\Omega_4 x_1}\right]\sin(\kappa x_2); \tag{16}$$

here $\kappa = 2\pi/\ell$ is the wavenumber of the bifurcated mode, $b_j$ ($j = 1,\dots,4$) are arbitrary coefficients and $\Omega_j$ ($j = 1,\dots,4$) represent the roots of the characteristic equation associated to Eq. (15),

$$\Omega_j^2 = \frac{1-2\xi+(-1)^j\Lambda}{1+k}, \quad \text{where} \quad \Lambda = \sqrt{k^2+4\xi(\xi-1)}. \tag{17}$$

Coefficients $\Omega_j$ satisfy $\Omega_3 = -\Omega_1$ and $\Omega_4 = -\Omega_2$. In particular, according to the sign of $\Lambda^2$, the $\Omega_j$ are:

- complex conjugate pairs within the elliptic complex regime (EC), $\Omega_2 = -\overline{\Omega}_1$, so that $\Omega_1$ and $\Omega_2$ share the same imaginary part;
- purely imaginary conjugate pairs in the elliptic imaginary regime (EI).

The incremental solution must satisfy the decay condition at infinity, so that only the exponents $\Omega_j$ with positive imaginary parts are to be retained. Denoting such roots by $\Omega_1$ and $\Omega_2$ under the assumption $\mathrm{Im}[\Omega_1] \geq 0$ and $\mathrm{Im}[\Omega_2] \geq 0$, the stream function becomes

$$\psi(x_1,x_2) = -\frac{1}{i\kappa}\left[b_1 e^{i\kappa\Omega_1 x_1} + b_2 e^{i\kappa\Omega_2 x_1}\right]\sin(\kappa x_2). \tag{18}$$

Upon substituting the expression for the stream function into Eqs. (14), the components of the incremental displacement are expressed in the form

$$\begin{aligned} v_1 &= i\left[b_1 e^{i\kappa\Omega_1 x_1} + b_2 e^{i\kappa\Omega_2 x_1}\right]\cos(\kappa x_2), \\ v_2 &= \left[b_1\Omega_1 e^{i\kappa\Omega_1 x_1} + b_2\Omega_2 e^{i\kappa\Omega_2 x_1}\right]\sin(\kappa x_2). \end{aligned} \tag{19}$$

Eventually, integration of Eqs. (13), with the stream function expressed in Eq. (18), provides the incremental mean in-plane stress as

$$\dot{\Pi} = -\mu\kappa\left[b_1(k-\Lambda)\Omega_1 e^{i\kappa\Omega_1 x_1} + b_2(k+\Lambda)\Omega_2 e^{i\kappa\Omega_2 x_1}\right]\cos(\kappa x_2), \tag{20}$$

so that the incremental nominal stresses in Eqs. (10) become

$$\begin{aligned} \frac{\dot{t}_{11}}{\mu\kappa\cos(\kappa x_2)} &= b_1\,(\eta-2\xi+\Lambda)\,\Omega_1 e^{i\kappa\Omega_1 x_1} + b_2\,(\eta-2\xi-\Lambda)\,\Omega_2 e^{i\kappa\Omega_2 x_1}, \\ \frac{\dot{t}_{22}}{\mu\kappa\cos(\kappa x_2)} &= -b_1\,(\eta-2\xi-\Lambda)\,\Omega_1 e^{i\kappa\Omega_1 x_1} - b_2\,(\eta-2\xi+\Lambda)\,\Omega_2 e^{i\kappa\Omega_2 x_1}, \\ \frac{\dot{t}_{12}}{i\mu\kappa\sin(\kappa x_2)} &= b_1\,(\eta-2\xi-\Lambda)\,e^{i\kappa\Omega_1 x_1} + b_2\,(\eta-2\xi+\Lambda)\,e^{i\kappa\Omega_2 x_1}, \\ \frac{\dot{t}_{21}}{i\mu\kappa\sin(\kappa x_2)} &= b_1\left[k-1+(1-\eta)\,\Omega_1^2\right]e^{i\kappa\Omega_1 x_1} \\ &\quad + b_2\left[k-1+(1-\eta)\,\Omega_2^2\right]e^{i\kappa\Omega_2 x_1}. \end{aligned} \tag{21}$$

## 3. Wrinkling analysis via the 'surface apparent rigidity' approach

Following Biot [1], the perturbing sinusoidal dead load defined in Eq. (8) is superimposed on the surface of the prestressed half-space. The incremental boundary conditions (9) are enforced on the surface $x_1 = 0$ where $p = -T_1$, thus setting the parameter $\alpha$ to $-(\eta+k)$ and 0 for pressure and dead load, respectively. The incremental displacement normal to the surface is

$$v_1(x_2) = v\cos(\zeta x_2), \tag{22}$$

where $v$ represents the amplitude. Therefore, the dimensionless 'apparent rigidity' of the free surface is defined according to Biot [1] as

$$\varphi = \frac{\sigma}{2\mu\, v\,\zeta}, \tag{23}$$

and turns out to vanish when a surface bifurcation is met. By connecting the amplitudes of the incremental displacement $v$ and the perturbing load $\sigma$, a novel expression for the surface apparent rigidity is derived,

$$\varphi(\alpha,\xi,k,\eta) = i\,\frac{\gamma_1^2\,\Omega_1 - \gamma_2^2\,\Omega_2}{4\Lambda}, \quad \gamma_j = \alpha+\eta-2\xi-(-1)^j\Lambda, \quad j = 1,2. \tag{24}$$

Interestingly, the surface rigidity, Eq. (24), is valid for any prestress state and turns out to depend on $\eta$ only through the coefficients $\gamma_j$. Therefore, enforcing the boundary condition associated with pressure/suction, $\alpha + \eta = -k$, the coefficients $\gamma_i$ become independent of $\eta$ and it follows that:

*for pressure/suction, the surface apparent rigidity $\varphi$ becomes independent of $\eta$.*

The onset of surface bifurcation is associated with the vanishing of the surface rigidity. Eq. (24) is specialized to the elliptic regimes (EC) and (EI) for both pressure and dead loading conditions as follows.

- (*EC*) for $1-\sqrt{1-k^2} < 2\xi < 1+\sqrt{1-k^2}$, as $i\Lambda < 0$, the roots $\Omega_j$ are complex numbers with non-vanishing real parts,

$$\Omega_1 = -\Gamma_1 + i\Gamma_2, \qquad \Omega_2 = -\overline{\Omega}_1 = \Gamma_1 + i\Gamma_2, \tag{25}$$

  where

$$\Gamma_j = \sqrt{\frac{\sqrt{1-k^2}-(-1)^j(1-2\xi)}{2(1+k)}}, \quad (j = 1,2). \tag{26}$$

- (*EI*) for $2\xi > 1+\sqrt{1-k^2}$, as $\Lambda^2 > 0$, while $\Omega_j^2 < 0$ ($j = 1,2$), purely imaginary roots $\Omega_j$ are found,

$$\Omega_1 = i\,\Delta_1, \qquad \Omega_2 = i\,\Delta_2, \tag{27}$$

  where $\Delta_j \in \mathbb{R}$.

### 3.1. Tensile and compressive wrinkling for neo-Hookean elasticity

The neo-Hookean case ($\xi = 1$) is especially relevant, as its simplicity makes the main bifurcation features easy to identify. The condition $\xi = 1$ marks the transition between the two elliptic regimes, yielding $\Lambda = |k|$, so that the two admissible roots become

$$\left.\begin{array}{l}\Omega_1\\ \Omega_2\end{array}\right\} = i\sqrt{\frac{1 \pm |k|}{1+k}}. \tag{28}$$

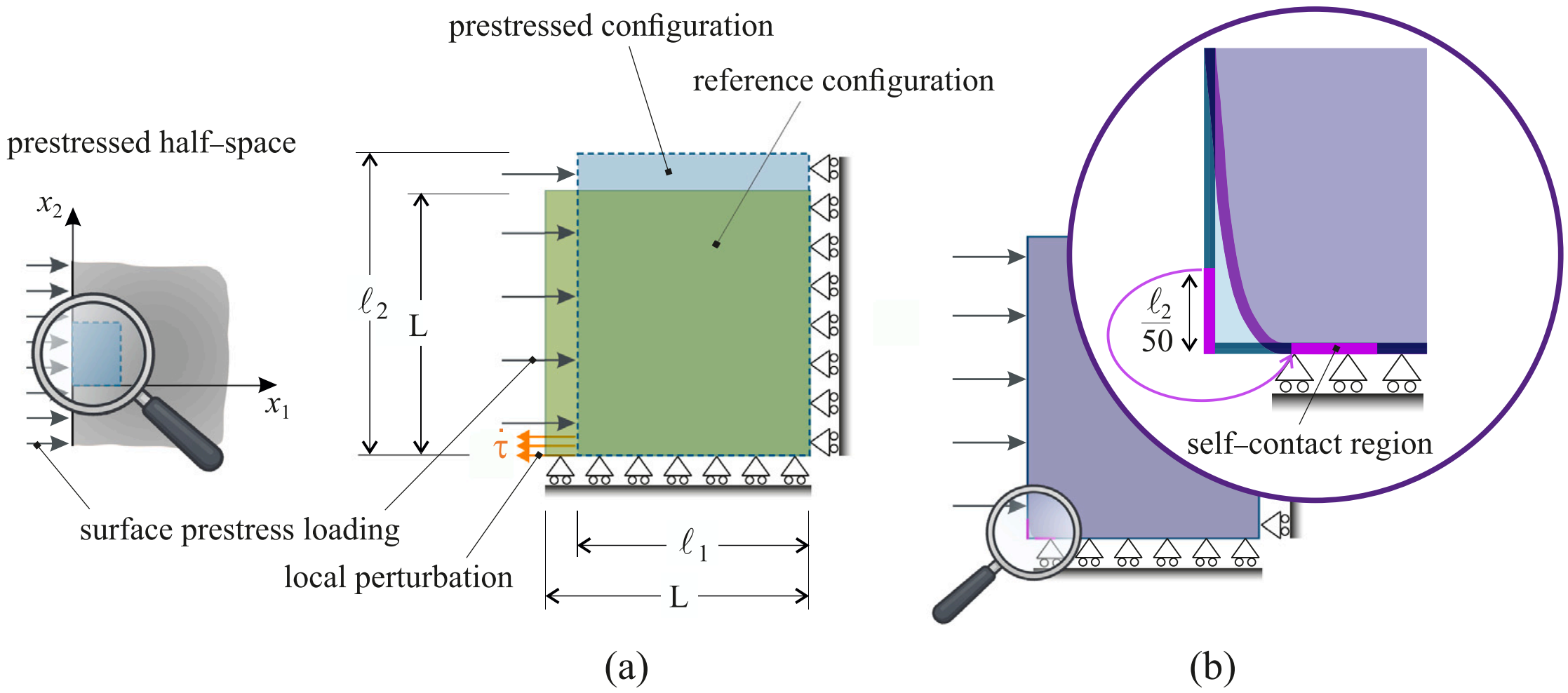

**Fig. 1.** Square domain near the free surface excised from the reference configuration of a half-space for Abaqus 2022 simulations. (a) The reference domain (shown green), has been chosen as a $L \times L$ square, and is prestressed in order to reach a desired $\ell_1 \times \ell_2$ rectangular configuration (shown blue). Wrinkling is analyzed through superposition of a perturbative incremental force $\dot{\tau}$, distributed along the 3 outermost elements on the boundary. (b) The post-critical distortion enforced on the prestressed configuration to analyze creasing instability. (For interpretation of the references to color in this figure legend, the reader is referred to the web version of this article.)

When $\xi = 1$ and a *uniaxial prestress normal to the surface* prevails, $\eta = k = T_1/(2\mu)$, the surface rigidity computed from Eq. (24) vanishes when

$$k^3 \pm 2(k^2 - 1) = 0, \tag{29}$$

where the sign '+ ' (the sign '−') applies for pressure load (for dead load). The solutions to Eqs. (29) can be written as

$$k = \frac{1}{3}\left(y\left[(-1)^{\frac{2}{1-a}} + (-1)^{-\frac{2}{1-a}}\frac{a^2}{y^2}\right] - a\right),$$

where $a = +2$ for pressure/suction, $a = -2$ for dead load, and

$$y = \sqrt[3]{(19/2)a + 3\sqrt{33}}.$$

The following real solutions for bifurcation are found.

- *Tensile* bifurcation for *pressure load*,

  $k = +0.8393, \quad T_1 = 1.6786\,\mu = 3.08738\,\mu_0, \qquad \text{at} \quad \lambda_1 = 1.8393;$

- *Compressive* bifurcation for *dead load*,

  $k = -0.8393, \quad T_1 = -1.6786\,\mu = -3.08738\,\mu_0, \quad \text{at} \quad \lambda_1 = 0.5437.$

The expression $\mu = \mu_0/2\left(\lambda_1^2 + \lambda_1^{-2}\right)$ has been used above, relating the current shear modulus $\mu$ to its counterpart in the undeformed configuration $\mu_0$. Note that both critical stretches correspond to $\mu = 1.839\,\mu_0$ and to the same logarithmic strain, differing only in sign, $\log \lambda_1 = \pm 0.60936$.

Biot [2] showed that wrinkling occurs when the half-space experiences a uniaxial compressive stress oriented parallel to the surface; however, the same condition is found also when the compressive prestress is oriented orthogonal to it, and is generated by a dead load.

*A numerical analysis of wrinkling* for neo-Hookean material confirms the theoretically predicted bifurcation conditions. The analysis, performed through Abaqus 2022 (standard version), represents a preliminary step toward the subsequent numerical investigation of creasing. The scale-free problem of an elastic half-space under uniform surface loading is modeled by ideally excising a square $L \times L$ planar region from the infinite domain in its unloaded configuration, sharing a portion of its free surface, Fig. 1(a). A pressure, or a dead load, is applied horizontally on the left side, while two orthogonal edges are constrained with rollers; all the surfaces are assumed to be frictionless. The desired state of prestress is enforced, so to attain, under a null vertical stress, the uniform stretches $\lambda_1 = \ell_1/L$ and $\lambda_2 = \ell_2/L$, which satisfy incompressibility, so that $\lambda_1 = 1/\lambda_2$ under plane strain assumption. In this condition, the neo-Hookean and Mooney–Rivlin materials prove to be equivalent, and are used here with the strain energy density $U = \mu_0\left(\lambda_1^2 + \lambda_2^2 - 2\right)/2$.

The numerical modeling is based on $500 \times 500$ hybrid plane strain 4-noded elements CPE4H. Once the prestressed configuration is obtained through a nonlinear analysis, the effects of a distributed dead load $\dot{\tau}$ superimposed in the normal direction over a small region of the surface (corresponding to the three outermost elements) are computed through a linear perturbation analysis. Numerical results are represented in Fig. 2, where the evolution of the maximum incremental displacement $\Delta v_1$ induced by $\dot{\tau}$, made dimensionless by multiplication with $\mu/(L\dot{\tau})$, is pictured with respect to the prestress parameter $k = T_1/(2\mu)$ (describing a uniaxial stress state). Wrinkling corresponds to a vanishing surface rigidity [1], namely to a blow-up of the incremental displacement. The graph shows this instability occurring under suction at $k = T_1/(2\mu) = 0.8393$. For a dead load, wrinkling arises only under compression, $k = T_1/(2\mu) = -0.8393$, with an additional bifurcation occurring in tension at the boundary of the elliptic domain, $k \to 1$, where a shear band parallel to the applied load is known to develop [4]. The two insets show the perturbation-induced deformation at the bifurcation prestress states, $k = \pm 0.839285$, representing the wrinkling onset modes.

### 3.2. Results on wrinkling under suction/pressure vs. dead load

The condition of vanishing surface rigidity is enforced via Eq. (24) for pressure/suction and dead load, with the latter used for comparison. Fig. 3 provides pictorial representations of the results in the biaxial prestress plane $T_1/\mu$– $T_2/\mu$ for neo-Hookean elasticity ($\xi = 1$) and anisotropic elasticity ($\xi = 0.25$), respectively in panels (*a*) and (*b*). In the figures, the (EI) and (EC) regimes are shaded green with different intensities.

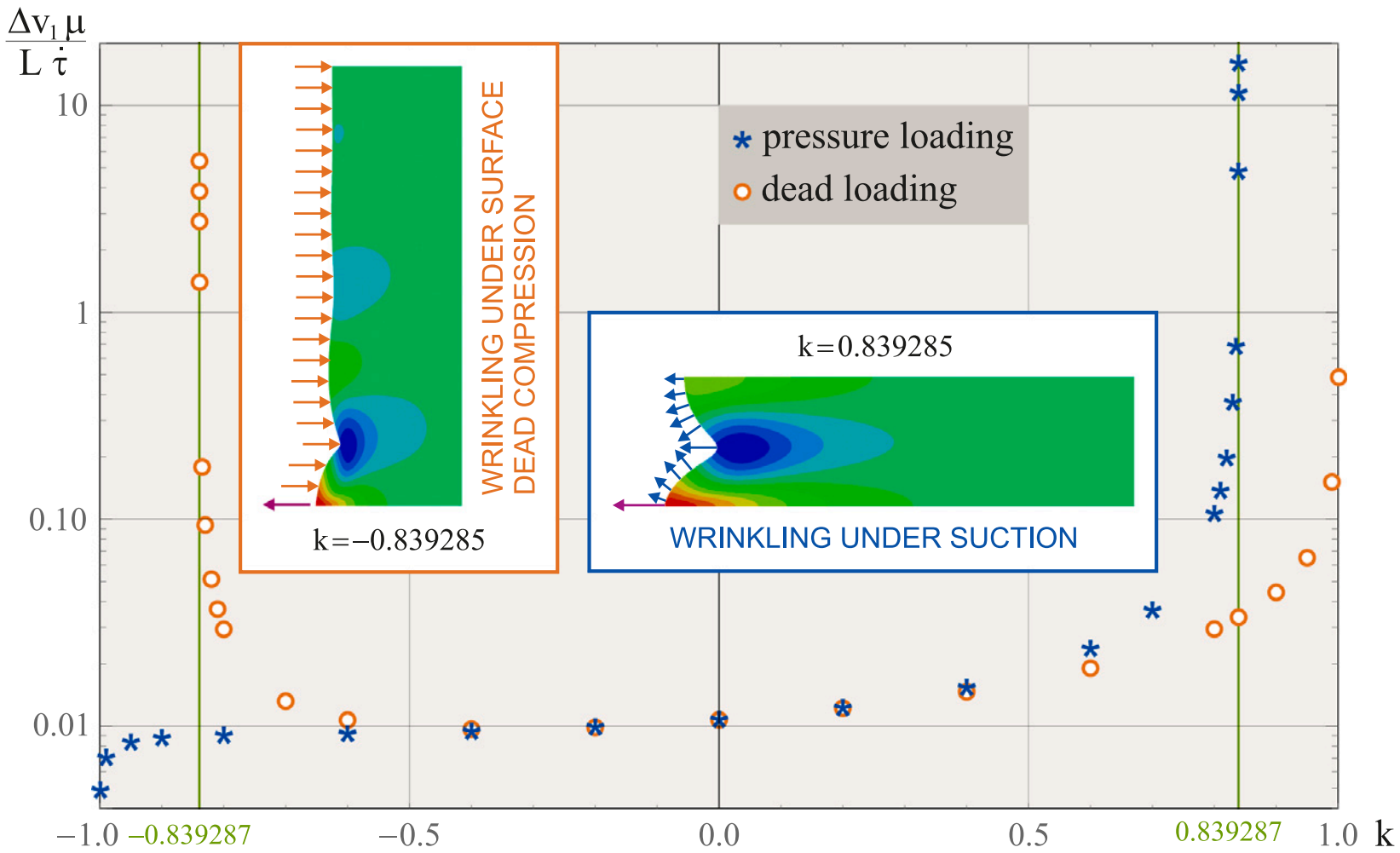

**Fig. 2.** Wrinkling is detected in a prestressed half-space (insets) subjected to a perturbing dead load $\dot{\tau}$ by a sharp loss of surface rigidity and the consequent blow-up of the incremental displacement $\Delta v_1$ (shown in nondimensional form on a logarithmic scale).

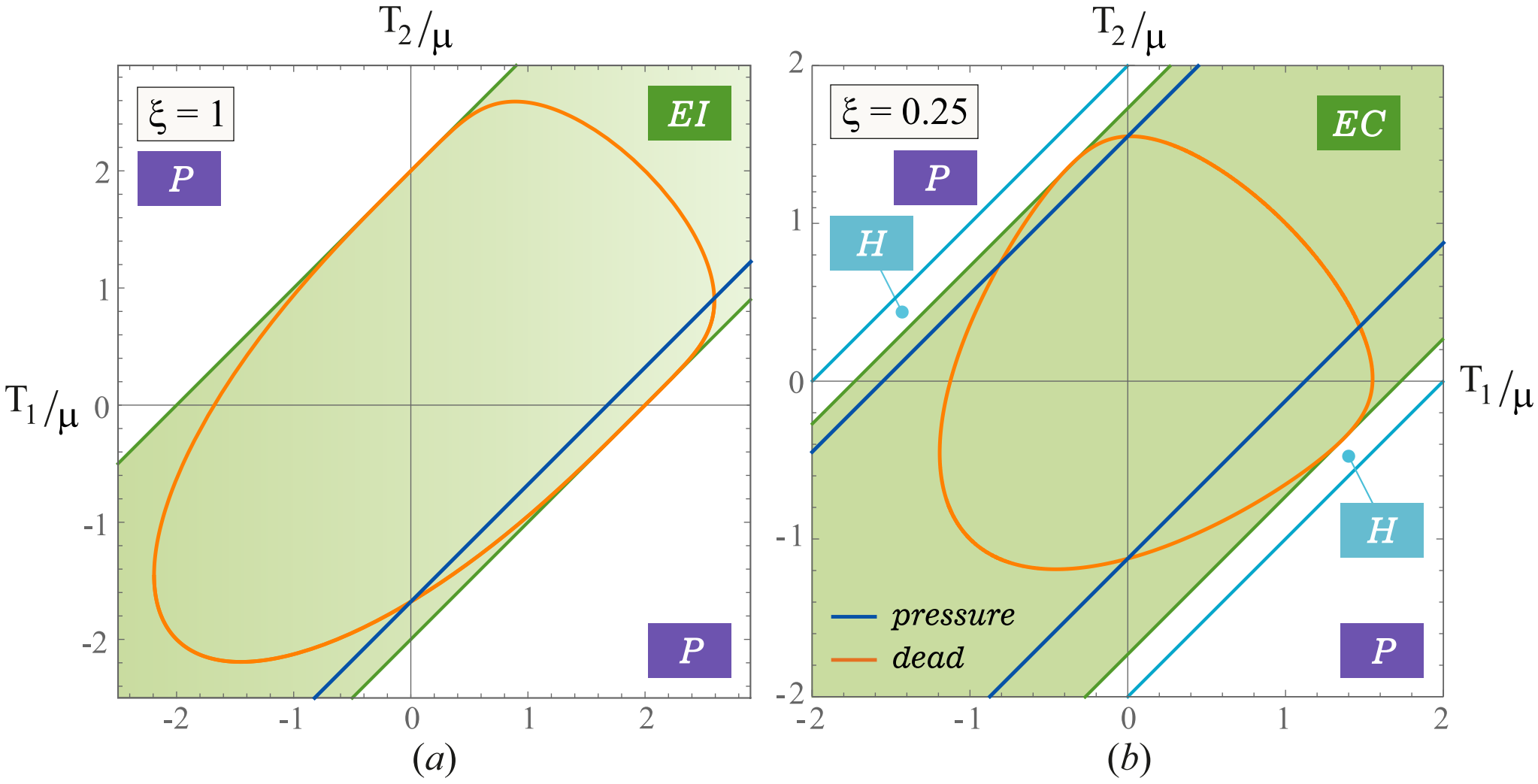

**Fig. 3.** Wrinkling conditions under surface pressure/suction (blue straight lines) in the biaxial prestress plane for (a) neo-Hookean elasticity ($\xi = 1$) and (b) anisotropic elasticity ($\xi = 0.25$). The corresponding instability condition for dead load is included for comparison as the orange closed curve. Under suction, wrinkling can occur under uniaxial stress and the no-wrinkling domain is open, whereas for dead load it is always closed. For $\xi < 0.5$, two wrinkling conditions under surface pressure/suction exist (blue straight, parallel lines; shown here for $\xi = 0.25$), allowing wrinkling under both uniaxial tension and compression. (For interpretation of the references to color in this figure legend, the reader is referred to the web version of this article.)

The condition for wrinkling under pressure or suction – independent of $\eta$ – is indicated in the figure on the left (on the right) by a blue straight line (by two blue straight lines), thereby precluding wrinkling to its left (inside the region comprised between the two lines). The orange curved lines, delimiting closed sets of prestress states, correspond to the attainment of the condition of wrinkling for dead load; consequently, all prestress states lying inside the set ensure uniqueness of the incremental response.

Note that the orange curves are symmetric with respect to an axis inclined at $\pi/4$ and are tangent to the (E) boundaries on both sides, being locally nearly straight. The symmetry shows that, under dead load, $T_1$ and $T_2$ play the same role in wrinkling, which is not the case for pressure/suction, where wrinkling requires a condition well inside (E) and parallel to its limit lines.

A special feature of pressure/suction, in contrast to dead load, is that the wrinkling locus in the prestress plane is represented by open – rather than closed – regions, so that the prestress can be increased up to infinity along a path inclined at $\pi/4$ without generating any bifurcation or (E) loss. This feature is also confirmed by the numerical analyses carried out within the elliptic regime for a pure compression and a loading $T_1 = T_2$, both significantly exceeding the wrinkling thresholds for dead load.

The neo-Hookean material exhibits coincident incremental shear moduli $\mu_* = \mu$; an investigation of more complex anisotropic behavior for $\xi \neq 1$ can easily be performed through Eqs. (24)–(26). Results from this investigation are reported in Fig. 3 on the right and can be summarized as follows.

- For $\xi > 1$, the features of the left figure remain valid; with increasing $\xi$, the wrinkling line for suction asymptotically approaches the right boundary between regimes (EI) and (P).
- For $0.5 < \xi < 1$, both regimes (EC) and (EI) develop, with the latter being bordered by the (P) regime. The line representing the conditions for wrinkling under suction lies within the (EC) or (EI) domain for $0.5 < \xi < 0.8$ or $0.8 < \xi < 1$, respectively.
- For $\xi \leq 0.5$, two conditions are identified for wrinkling under pressure/suction, both lying well-inside the regime (EC), whereas regime (EI) no longer appears. These conditions are represented by two (blue) straight lines parallel to each other and to the boundaries between regimes (EC) and (H), and between (H) and (P). In this case, wrinkling is foreseen for both uniaxial tension and compression.

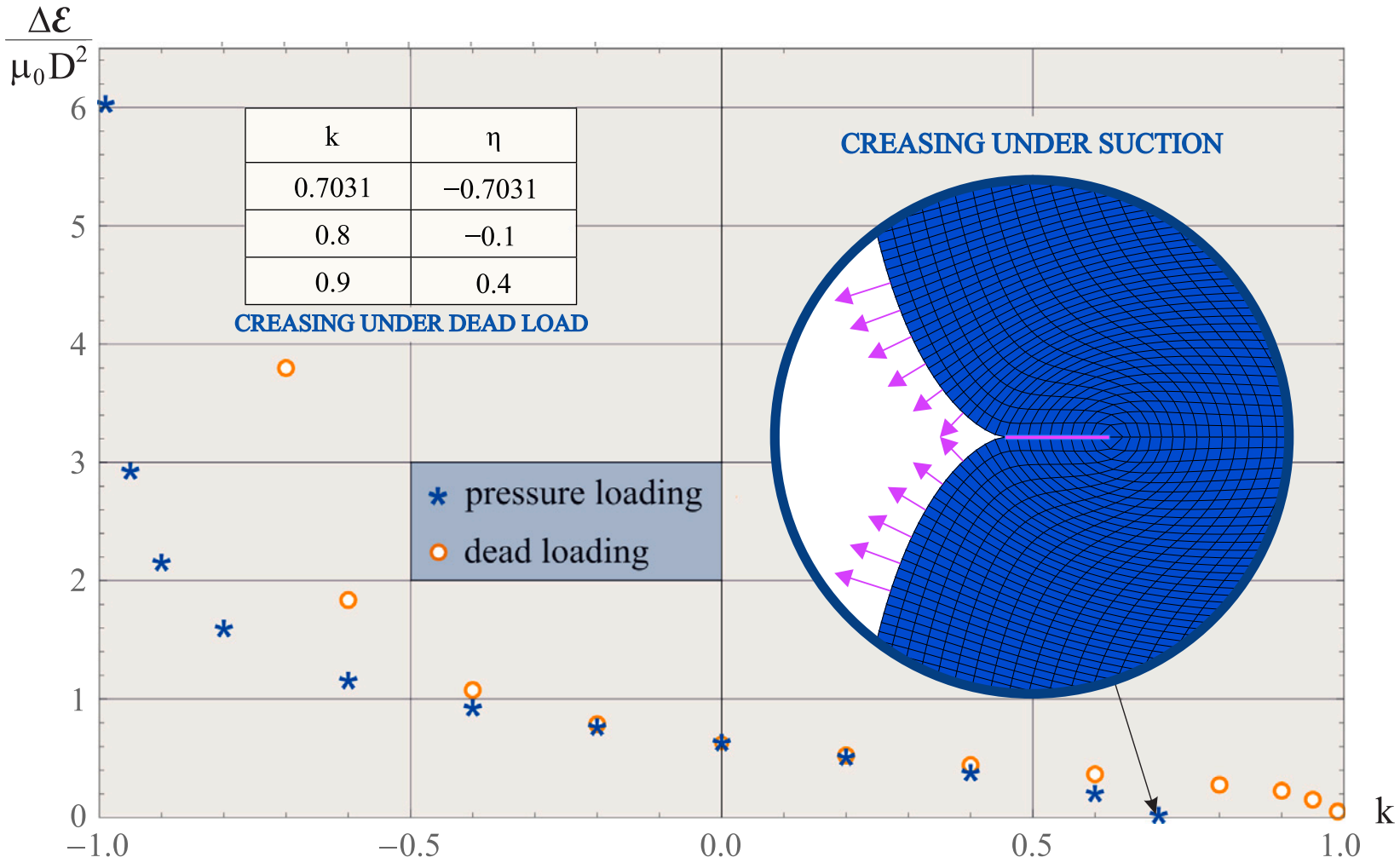

**Fig. 4.** Surface creasing of an elastic half-space subject to pressure/suction or dead load on its surface. Creasing may occur for suction but also for pressure when $k = 0.7031$, while it is excluded for dead load, where the condition of ellipticity loss prevails, $k = 1$. Creasing is detected when the energy difference $\Delta\mathcal{E}$ between the uniform prestressed state and the creased state vanishes as a function of the prestress $k$. Under pressure/suction, creasing is found to be independent of $\eta$, so that the results reported in the figure are representative of all possible prestress states. Conversely, creasing under dead loading is found to be dependent on $\eta$, as shown in the inset, thus the reported results refer only to a uniaxial state of prestress $k = T_1/(2\mu)$. (For interpretation of the references to color in this figure legend, the reader is referred to the web version of this article.)

The occurrence of wrinkling under suction for general anisotropic elasticity suggests that the instability is not merely a consequence of the constitutive assumptions, but rather reflects a more general effect related to the boundary conditions. This interpretation is consistent with the failure of the complementarity condition discussed in [3].

## 4. Creasing instability for a half-space under pressure/suction or dead load

For a neo-Hookean half-space under homogeneous compression parallel to the free surface, the critical stretch for *wrinkling*, namely, $\lambda_w = 0.5437$, is smaller than the experimental threshold for *creasing*, $\lambda_c = 0.65$ [9]. As a highly localized self-contacting fold, creasing cannot be described by Biot's linear perturbation analysis and instead requires a nonlinear analysis with a suitable perturbation, such as a fold [15] or a concentrated force normal to the surface [20]. The occurrence of creasing instability on the surface of a half-space *loaded by a distribution of orthogonal forces* represents a still unexplored field. In this context, the peculiarities found for wrinkling suggest that analogous features are in principle observable for creasing; this is the aim of the present section, where the instability is investigated through an adaptation of the approach pointed out in [15].

The same square finite domain and preloading procedure described in Section 3.1 for the wrinkling analysis of neo-Hookean material are adopted here to numerically investigate stability with respect to creasing. After a convergence analysis to assess the sensitivity of the solution to element size and type, domain shape, and contact region geometry, the finite element technique illustrated in [15] is adapted to the present context, assuming a small portion of the free surface (having initial length $D = L/50$, and undergoing the same pressure – or dead load – applied on the whole surface), as pictured in Fig. 1(b). Its nodal points, initially aligned along the vertical axis $x_2$, are prescribed a vertical displacement in order to realign them onto the horizontal axis $x_1$, a computational expedient to enforce the frictionless self-contact on the free surface associated with a crease development. Although the horizontal displacement of the nodes remains unconstrained during this deformation (and is computed numerically), their new horizontal positions preserve their initial order. Starting from a stable prestressed state, the achievement of the distorted configuration requires a positive increment of the elastic energy $\Delta\mathcal{E}$, computed by Abaqus through a nonlinear analysis. The condition $\Delta\mathcal{E} = 0$ identifies the onset of creasing, which becomes energetically favorable.

The evolution of $\Delta\mathcal{E}$ as a function of the prestress is shown in Fig. 4, for both pressure (data points marked by blue asterisks) and dead load (orange circles). In agreement with the results obtained for wrinkling, the following conclusions can be reached.

- The results for pressure or suction are found to be independent of the prestress parameter $\eta$, so that Fig. 4 refers to all possible prestress states. Creasing becomes possible at $k = 0.7031$, corresponding to a horizontal stretch $\lambda_1 = 1.5476$; this has a vertical counterpart $\lambda_2 = 0.646$, in agreement with the results obtained in [15,20] for a half-space having a free surface and loaded parallel to it. Note that the critical value $k = 0.7031$ can be attained under infinitely many prestress combinations, including surface pressure/suction and normal load. A relevant example is the uniaxial stress state induced by suction, $T_1/\mu = 1.4062$.
- For a dead surface load, the plotted results correspond exclusively to a uniaxial state of prestress, $k = \eta = T_1/(2\mu)$: creasing is only possible under tension at the ellipticity loss (which is not achievable for a neo-Hookean material), a feature related to the emergence of a horizontal shear band. More in general, creasing is found to depend on $\eta$ when a dead load is applied on the surface. This dependence is illustrated for a few values of $\eta$ in the inset of Fig. 4. Further investigation of dead load falls beyond the scope of this article.
- When the state of prestress is uniaxial and orthogonal to the surface, for a neo-Hookean material, the following features, not previously reported, are highlighted: (i) the suppression of wrinkling and creasing under applied pressure; (ii) the occurrence of wrinkling, together with the disappearance of creasing, under compressive dead load. The latter finding is noteworthy, as it may provide a procedure for inducing wrinkling while preventing creasing.

In closure, note that several authors suggest that inside the prestress interval between creasing $k = 0.7031$ and wrinkling $k = 0.8393$, the surface becomes metastable [11,12,20]. Our loading conditions differ from those previously analyzed; however, the interpretation may still be valid.

## 5. Conclusions

Creasing and wrinkling in an elastic half-space have been investigated under surface suction or pressure. For neo-Hookean elasticity, both instabilities are found to be excluded when an applied pressure generates a uniaxial stress state. However, it is surprising that they may arise under suction. While they represent tensile instabilities, their occurrence is similar to that observed when a compression is applied parallel to the otherwise free surface. Moreover, the stability domains corresponding to creasing and wrinkling when a pressure or suction is applied were found to be open; therefore, in principle, a loading scheme is determined that allows the material to reach an unbounded stress without meeting any instability. Suction and pressure play a central role as boundary conditions in many technological and biomechanical settings, where surface loading induced by fluid or environmental interactions can critically influence stability and deformation behavior.

## CRediT authorship contribution statement

**Roberta Springhetti:** Writing – review & editing, Writing – original draft, Visualization, Validation, Supervision, Software, Resources, Methodology, Investigation, Formal analysis, Conceptualization. **Davide Bigoni:** Writing – review & editing, Writing – original draft, Visualization, Validation, Supervision, Resources, Project administration, Methodology, Investigation, Funding acquisition, Formal analysis, Conceptualization.

## Declaration of competing interest

The authors declare that they have no known competing financial interests or personal relationships that could have appeared to influence the work reported in this paper.

## Acknowledgments

The authors acknowledge funding from the European Research Council (ERC) under the European Union's Horizon Europe research and innovation programme, Grant agreement No. ERC-ADG-2021-10 1052956-BEYOND.

## Data availability

No data was used for the research described in the article.

## References

[1] M.A. Biot, Surface instability of rubber in compression, Appl. Sci. Res. Sect. A 12 (1963) 168–182.
[2] M.A. Biot, Mechanics of Incremental Deformations, John Wiley & Sons, Inc., 1965.
[3] A. Benallal, R. Billardon, G. Geymonat, Bifurcation and localization in rate-independent materials. Some general considerations, in: Q.S. Nguyen (Ed.), Bifurcation and Stability of Dissipative Systems, Springer Vienna, Vienna, 1993, pp. 1–44.
[4] D. Bigoni, Nonlinear Solid Mechanics: Bifurcation Theory and Material Instability, Cambridge University Press, 2012.
[5] M.A. Dowaikh, R.W. Ogden, On surface waves and deformations in a pre-stressed incompressible elastic solid, IMA J. Appl. Math. 44 (3) (1990) 261–284.
[6] M. Hayes, R.S. Rivlin, Surface waves in deformed elastic materials, Arch. Ration. Mech. Anal. 8 (1961) 358–380.
[7] R. Hill, J. Hutchinson, Bifurcation phenomena in the plane tension test, J. Mech. Phys. Solids 23 (4) (1975) 239–264.
[8] T. Tanaka, S.T. Sun, Y. Hirokawa, S. Katayama, J. Kucera, Y. Hirose, T. Amiya, Mechanical instability of gels at the phase transition, Nature 325 (1987) 796–798.
[9] A.N. Gent, I.S. Cho, Surface instabilities in compressed or bent rubber blocks, Rubber Chem. Technol. 72 (1999) 253–262.
[10] E. Hohlfeld, L. Mahadevan, Unfolding the sulcus, Phys. Rev. Lett. 106 (2011) 105702.
[11] Y. Cao, J. Hutchinson, From wrinkles to creases in elastomers: The instability and imperfection-sensitivity of wrinkling, R. Soc. Lond. Proc. Ser. A 468 (2011) 94–115.
[12] P. Ciarletta, Matched asymptotic solution for crease nucleation in soft solids, Nat. Commun. 9 (2018) 496.
[13] P. Ciarletta, L. Truskinovsky, Soft nucleation of an elastic crease, Phys. Rev. Lett. 122 (2019) 248001.
[14] S.S. Pandurangi, A. Akerson, R.S. Elliott, T.J. Healey, N. Triantafyllidis, Nucleation of creases and folds in hyperelastic solids is not a local bifurcation, J. Mech. Phys. Solids 160 (2022) 104749.
[15] W. Hong, X. Zhao, Z. Suo, Formation of creases on the surfaces of elastomers and gels, Appl. Phys. Lett. 95 (11) (2009) 111901.
[16] M. Rossi, G. Cicconofri, A. Beran, G. Noselli, A. DeSimone, Kinematics of flagellar swimming in euglena gracilis: Helical trajectories and flagellar shapes, Proc. Natl. Acad. Sci. 114 (50) (2017) 13085–13090.
[17] S. Sareh, J.M. Rossiter, A.T. Conn, K. Drescher, R.E. Goldstein, Swimming like algae: Biomimetic soft artificial cilia, J. R. Soc. Interface 10 (2013).
[18] G. Cicconofri, V. Damioli, G. Noselli, Nonreciprocal oscillations of polyelectrolyte gel filaments subject to a steady and uniform electric field, J. Mech. Phys. Solids 173 (2023) 105225.
[19] M. Ben Amar, Wrinkles, creases, and cusps in growing soft matter, Rev. Modern Phys. 97 (2025) 015004.
[20] P. Yang, Y. Fang, Y. Yuan, S. Meng, Z. Nan, H. Xu, H. Imtiaz, B. Liu, H. Gao, A perturbation force based approach to creasing instability in soft materials under general loading conditions, J. Mech. Phys. Solids 151 (2021) 104401.